\documentclass[useAMS,usenatbib]{mn2e}

\usepackage[psamsfonts]{amssymb}
\usepackage[dvips]{graphicx}
\usepackage{amsmath,alltt}
\usepackage{multirow}
\usepackage{rotating}
\usepackage{lscape}

\title[Primordial binaries in globular clusters]{The state of globular clusters at birth II: primordial binaries}
\author[Leigh et al.]{Nathan Leigh$^{1,2}$, Mirek Giersz$^{3}$, Michael Marks$^{4}$, Jeremy J. Webb$^{5}$, 
Arkadiusz Hypki$^{3,6}$, 
\newauthor 
Craig Heinke$^{1}$, Pavel Kroupa$^{4}$, Alison Sills$^{5}$
\thanks{E-mail: nleigh@ualberta.ca (NL); mig@camk.edu.pl (MG); mmarks@astro.uni-bonn.de (MM); webbjj@mcmaster.ca (JW); ahypki@camk.edu.pl (AH); heinke@ualberta.ca (CH); pavel@astro.uni-bonn.de (PK); asills@mcmaster.ca (AS)}\\
$^{1}$Department of Physics, University of Alberta, CCIS 4-183, Edmonton, AB T6G 2E1, Canada \\
$^{2}$Department of Astrophysics, American Museum of Natural History, Central Park West and 79th Street, New York, NY 10024 \\
$^{3}$Nicolaus Copernicus Astronomical Centre, Polish Academy of Sciences, ul. Bartycka 18, 00-716 Warsaw, Poland \\
$^{4}$Helmholtz-Institut f\"ur Strahlen- und Kernphysik, Nussallee 14-16, D-53115, Bonn, Germany \\
$^{5}$McMaster University, Department of Physics and Astronomy, 1280 Main St. W.,
Hamilton, Ontario, Canada, L8S 4M1 \\
$^{6}$Leiden Observatory, Leiden University, P.O. Box 9513, 2300 RA Leiden, The Netherlands}
\begin{document}

\pagerange{\pageref{firstpage}--\pageref{lastpage}} \pubyear{2011}

\maketitle

\label{firstpage}

\begin{abstract}


In this paper, we constrain the properties of primordial binary populations 
in Galactic globular clusters.  Using the MOCCA Monte Carlo code for cluster 
evolution, our simulations cover three decades in present-day total cluster mass.  Our 
results are compared to the observations of Milone et al. (2012) using the photometric binary 
populations as proxies 
for the true underlying distributions, in order to test the hypothesis that the data are 
consistent with an universal initial binary fraction near unity and the binary orbital parameter 
distributions of Kroupa (1995).  With the exception of a few possible outliers, we find that the 
data are to first-order consistent with the 
universality hypothesis.  Specifically, the present-day binary fractions \textit{inside} the half-mass 
radius can be reproduced assuming either high initial binary fractions near unity with a dominant 
soft binary component as in the Kroupa distribution combined 
with high initial densities (10$^4$-10$^6$ M$_{\odot}$ pc$^{-3}$), or low initial binary 
fractions ($\sim$ 5-10\%) with a dominant hard binary component combined with moderate initial 
densities near their present-day values (10$^2$-10$^3$ M$_{\odot}$ pc$^{-3}$).  
This apparent degeneracy can potentially be broken using the binary fractions outside the half-mass 
radius - \textit{only high initial binary fractions with a significant soft component combined 
with high initial densities can contribute to 
reproducing the observed anti-correlation between the binary fractions outside the half-mass radius 
and the total cluster mass.}  
We further illustrate using the simulated present-day binary orbital parameter distributions and the 
technique first introduced in Leigh et al. (2012) that the relative fractions of hard and soft 
binaries can be used to further constrain both the initial cluster 
density and the initial mass-density relation.  Our results favour an initial mass-density 
relation of the form r$_{\rm h} \propto$ M$_{\rm clus}^{\alpha}$ with $\alpha <$ 1/3, corresponding 
to an initial correlation between cluster mass and density.
\end{abstract}

\begin{keywords}
binaries: general -- stars: kinematics and dynamics -- globular clusters: general -- stars: formation --
methods: numerical.
\end{keywords}

\section{Introduction} \label{intro}

The origin of globular clusters (GCs) remains an open question.  The initial 
conditions present during and after the gas-embedded phase are poorly 
constrained.  For instance, were GCs formed during the monolithic collapse 
of a single massive giant molecular cloud, or are they the merged remnants of 
many lower mass sub-clumps and/or filamentary structures?  Recent evidence suggests that 
globular clusters underwent 
prolonged star formation early on in their lifetimes (see \citet{gratton12} for a recent
review).  That is, gas could have been present in significant quantities for the first 
$\sim 10^8$ years, 
albeit perhaps intermittently \citep{conroy11,conroy12}.  This evidence comes
in the form of multiple stellar populations identified in the colour-magnitude
diagram \citep[e.g.][]{piotto07}, as well as curious abundance anomalies that cannot be
explained by a single burst of star formation \citep[e.g.][]{osborn71,gratton01}.  On the 
other hand, a prolonged gas-embedded phase seems to conflict with the recent findings 
of \citet{bastian14}, who 
report a lack of gas and dust in young massive clusters in the LMC and SMC, thought
to be (possible) analogs of primordial Galactic GCs.  In general, the presence 
of multiple populations in Galactic GCs is indicative of a complex formation history.  For 
example, massive clusters could re-accrete gas after an 
initial burst of star formation, allowing for a new generation of stars to form \citep{pflamm07}, 
or they might capture stars from the field in significant numbers if they are still forming 
in their natal molecular cloud \citep{pflamm09,fellhauer06}.  


In Paper I of this series \citep{leigh13c}, we constrained the origin of the observed dependences
of the present-day mass function (MF) slope and concentration parameter on the total
cluster mass.  
By focusing on the role of dynamics in modifying these parameters over $\sim 12$
Gyr of evolution, we showed that the present-day global stellar MF variations
are consistent with an universal initial MF modified by two-body relaxation and
Galactic tides, but some initial correlation between the concentration parameter
and the total cluster mass is needed to reproduce the present-day observed relations (a 
similar result was shown by \citet{marks10} to explain the present-day relation between 
total cluster mass and concentration, but invoking primordial gas expulsion).  We
further showed that the present-day distributions of core binary fractions can
be reproduced from an universal initial binary fraction of 10\%, assuming an
orbital period distribution that is flat in the logarithm of the semi-major axis
combined with an appropriate choice of the initial cluster mass-radius relation.  
Importantly, however, this did not preclude other combinations of the initial
binary fraction and orbital parameter distributions (i.e. the semi-major axis, 
mass ratio and eccentricity distributions).  

Among the many uncertainties pertaining to the origins of GCs are the properties of their 
primordial binary populations.  Much more is known about 
the \textit{present-day} binaries in these clusters, identified as 
photometric binaries above the main-sequence in the colour-magnitude 
diagram \citep[e.g.][]{milone12}, or at higher energies as exotic 
objects like low-mass x-ray binaries (LMXBs) \citep[e.g.][]{hut91}, millisecond 
pulsars (MSPs) \citep[e.g.][]{verbunt89} and cataclysmic variables (CVs) 
\citep[e.g.][]{pooley06,cohn10}.  Blue straggler (BS) formation as well is thought 
to involve binary stars, whether it be due to mass-transfer within a binary, 
collisions during encounters involving binaries or even some triple-based 
mechanism \citep[e.g.][]{leigh11a,geller13} such as Kozai-induced mergers 
\citep[e.g.][]{perets09,naoz14}.  Despite the importance of these populations for 
understanding a number of astrophysical processes, little is known about 
their progenitor populations and the conditions from which they 
evolved.

The principal physical mechanisms driving the time evolution of the binary orbital parameter 
distributions in dense star clusters are two-body relaxation and direct 
encounters involving binaries.\footnote{We note that in the binary-burning phase of initially 
binary-dominated clusters, the initial crossing time is of crucial importance.}  Two-body relaxation 
drives mass segregation in 
clusters, causing binaries to drift in toward the cluster centre.  This is 
due to the larger masses of binaries relative to single stars.  That is, the tendency for
clusters to evolve toward, on average, a state of energy equipartition (or, more accurately,
energy equilibrium, although technically this idealized state is never fully reached in
real clusters) results in a reduction in
the speeds of the more massive binaries.  Thus, two-body relaxation causes the inward
radial migration of binaries in clusters, with the most massive binaries having the
shortest mass segregation timescales \citep{vishniac78}.  Hence, this process does not directly 
impact the (global)
binary orbital parameter distributions.  Indirectly, however, mass segregation delivers
binaries to the central cluster regions where the density is highest, and encounters typically
occur on shorter timescales.\footnote{Mass segregation can also cause the average object mass
in the core to increase, which can in turn affect the hard-soft boundary.}  Binary encounters
have a more direct effect on the binary orbital parameter distributions via exchanges,
ionizations, hardening, softening, etc. \citep[e.g.][]{marks12,marks14}.  For example, exchanges
involve an interloping single star (or binary) that undergoes
a resonant gravitational interaction with the components of a binary and, in the end, can cause 
one of the original binary members to be ejected \citep{hills75,sigurdsson93}.  To summarize, mass 
segregation delivers binaries to the core, where they are dynamically processed 
due to binary encounters (ignoring the possible binary burning phase in the evolution of young and
dense clusters when binaries are destroyed on a crossing time \citep{marks11}).

The time evolution of the binary semi-major axis distribution in dense clusters can
be reasonably well summarized using one simple rule, called the Hills-Heggie Law
\citep{hills75,heggie75}.  This relies on the
concept of a ``hard-soft'' boundary, or the semi-major axis a$_{\rm HS}$ for which the binary
orbital energy is equal to the average single star kinetic energy.  That is:
\begin{equation}
\label{eqn:hard-soft}
a_{\rm HS} = \frac{G\bar{m}}{\sigma^2},
\end{equation}
where $\bar{m}$ is the average single star mass and $\sigma$ is the root-mean-square
(rms) velocity.  Here,
an average binary is defined such that both components have masses equal to the average
single star mass, and $\sigma$ is calculated using Equation 4-80b of \citet{binney87} (hence
it corresponds to the rms velocity at the half-light radius).
Binaries with (absolute) orbital energies greater than that corresponding to the hard-soft
boundary are called ``hard'', and typically experience a reduction in their semi-major axes
upon undergoing encounters with single stars.  ``Soft'' binaries, on the other hand,
have orbital energies less than the average single star kinetic energy, and become even
softer post-encounter as their semi-major axes typically increase.  The softest binaries
are easily ionized during encounters with single stars.  Thus, the net effect of binary
encounters is that hard binaries become harder, and soft binaries become softer.\footnote{This
statement relies on averages.  In reality, a binary that would be classified as hard
according to Equation~\ref{eqn:hard-soft} can actually be soft during a particularly
energetic encounter.  Or, binaries classified as soft by Equation~\ref{eqn:hard-soft} can be 
hard if the component masses are larger than the average stellar mass.}  This is the Hills-Heggie Law.

Recently, \citet{milone12} performed a photometric study of the 
main-sequence binary populations in a sample of 59 GCs.  The authors confirmed a
previously reported \citep{sollima07} anti-correlation between the binary
fraction and the total cluster mass.  \citet{sollima08} previously showed that 
this trend can arise naturally assuming an universal initial binary 
fraction.  In this scenario, the anti-correlation appears due to the disruption of 
soft binaries in the cluster core, combined with the evaporation of single stars from 
the cluster outskirts \citep[e.g.][]{duchene99,fregeau09,parker09}.  The efficiency of 
the former process should increase with increasing cluster mass and density \citep{marks11}, 
whereas the efficiency of the latter process is driven by two-body relaxation and should 
increase with decreasing cluster mass and density 
(i.e. t$_{\rm rel} \propto$ M$^{0.5}$r$_{\rm h}^{1.5}$).  
This contributes to high binary fractions in low-mass cluster cores, and low
binary fractions in high-mass cores.  

Even more recently, 
\citet{geller13} showed using $N$-body simulations that the radial dependence of 
the binary frequency is quite sensitive to the dynamical 
age of a cluster.  Dynamics initially destroys soft binaries 
in the core, giving rise to a binary frequency that rises toward the cluster 
outskirts.  After approximately one half-mass relaxation time in clusters with an 
initial mass 10$^4$-10$^5$ M$_{\odot}$, a bimodal radial 
binary frequency develops since binaries just outside the core have now drifted 
in due to mass segregation.  The minimum in binary frequency extends radially 
outward as time goes on, finally giving way to a binary fraction that decreases 
with increasing clustercentric distance all the way to the tidal radius after 
$\sim 4-6$ (initial) half-mass relaxation times.  Such a minimum in the binary 
fraction is clearly visible in NGC 5272 (M3), with a 6 Gyr half-mass relaxation 
time, around the half-mass radius \citep{milone12}.  The (incomplete) data of the 
comparable cluster NGC 5024 is suggestive of a similar minimum, and a binary 
fraction minimum at 1-2 half-mass radii is visible in the clusters NGC 6584 and 
6934, which have relaxation times of 1 Gyr.  An analogous bimodality has 
been observed in BS populations, and it is likely that mass segregation is 
again responsible for this (possibly transient; Hypki, A., 2014, PhD thesis) 
feature \citep{ferraro12}.  


In this paper, the second of the series, we delve further into the issue of
the initial binary properties in globular clusters.  In particular, we explore
whether or not the observed
present-day distribution of Galactic GC binary fractions can be reproduced
assuming an universal initial binary fraction of 95\% combined with the binary orbital
parameter distributions taken from \citet{kroupa95} (i.e. with a significant fraction of soft
binaries), which are deduced from present-day star-forming regions in the Milky Way disk and its 
field population.  We also explore different initial mass-density relations, but all of these 
correspond to relatively high initial densities ($\sim$ 10$^4$-10$^6$ M$_{\odot}$ pc$^{-3}$).  
To do this, we generate a series of simulations for GC evolution performed using the MOCCA
code \citep{giersz13}, and compare the results to the presently observed binary fractions
of \citet{milone12}.  

Importantly, there may never be an instant in time when the initial binary distribution functions are 
fully established.  In a dense star-forming environment, binaries form but can be dissociated into 
individual 
stars before other binaries form.  The binary population is constantly changing due to break-ups and 
mergers.  However, mathematically convenient initial binary distribution functions can be deduced from 
young stellar populations.  Crucially, the prior dynamical processing of the normalizing population 
must be properly accounted for.  This was first done by \citet{kroupa95} using constraints given by old 
populations.  The resulting mathematical formalism can be viewed as that which an ideal 
population would populate if it could.  For example, in an extreme star-burst that forms a very dense 
massive cluster, wide binaries cannot really form.  In this case, a binary fraction of 100\% seems 
unphysical.  But, we can model this using our derived mathematical description of primordial binary 
populations, since they become naturally truncated by the 
dynamical configuration of the cluster. Thus, the same initial binary distribution functions can be 
adopted, subject to the local cluster conditions.  These are taken care of naturally in any $N$-body 
or Monte Carlo model for GC evolution, leading to some slight cooling of the initial cluster 
configuration due to the disruption of wide binaries, although this typically represents at most 
a small fraction of the total energy budget of the cluster \citep{leigh13c}.

In Section~\ref{models}, we present our Monte Carlo models for GC evolution performed 
using MOCCA, along with our chosen initial conditions.  In Section~\ref{results}, we present 
our results and compare them to the observations.  We further describe 
our method, adapted from \citet{leigh12}, for extrapolating from only a handful of models 
convenient equations for the predicted present-day distributions of orbital 
energies for a range of present-day total cluster masses (spanning three orders of magnitude).  
This allows us to efficiently study and compare the evolution of our model binary populations 
in energy-space, as a function of different initial cluster densities.  Finally, in 
Section~\ref{discussion}, 
we discuss the implications of our results for primordial binaries in Milky Way GCs.

\section{Models} \label{models}

In this section, we describe the Monte Carlo code called MOCCA used 
to simulate the cluster evolution, and describe our choice of initial conditions.  
MOCCA is ideal for studies of globular cluster evolution spanning a range of initial conditions, 
given its fast and robust coverage of the relevant parameter space \citep[e.g.][]{giersz13}. 

\subsection{Monte Carlo models:  MOCCA} \label{mocca}

We use the MOCCA code to produce all of our simulated
clusters.  It combines the Monte Carlo technique for cluster evolution \citep{henon71} 
with the \textsc{FEWBODY} code \citep{fregeau04} to perform numerical
scattering experiments of small-number gravitational interactions.  The code 
relies on analytic formulae for stellar evolution taken from \citet{hurley00}, and
performs binary evolution calculations using the BSE code \citep{hurley02}.  
The MOCCA simulations are 
performed on a PSK2 cluster at the Nicolaus Copernicus Astronomical Centre in
Poland.  Each simulation is run on one CPU, and the cluster is based on AMD
Opteron processors with 64-bit architecture (2-2.4 GHz).  For further 
information about the MOCCA code, see \citet{hypki13}, \citet{giersz13} and 
\citet{leigh13c}.

\subsection{Initial conditions} \label{initial}

For every choice of Galactocentric radius and initial cluster 
structure, we run models having a total cluster mass of 10$^4$, 5 $\times$ 10$^4$, 
10$^5$, 5 $\times$ 10$^5$ and 8 $\times$ 10$^5$ M$_{\odot}$ initially.  
All models begin with a global initial binary fraction f$_{\rm b} =$ 95\%,\footnote{We restrict 
ourselves to a binary fraction slightly less than unity to avoid computational problems that 
arise in MOCCA if the number of single stars is initially zero.} and we do not assume any 
primordial binary (or stellar) mass segregation in our models.  The initial 
binary orbital parameter distributions are taken from Equation 46 of \citet{kroupa13}.  Therefore, 
initial eccentricities follow a thermal distribution f$_{\rm e}$(e) $=$ 2e.  
The initial binary mass function is derived from random pairing in the mass range 
0.08 to 5 M$_{\odot}$, and favouring mass ratios near unity above 5 M$_{\odot}$.  
The initial period distribution function is:
\begin{equation}
\label{eqn:period}
f_{\rm P} = {\eta}\frac{\log_{\rm 10}P - \log_{\rm 10}P_{\rm min}}{\delta + (\log_{\rm 10}P - \log_{\rm 10}P_{\rm min})^2},
\end{equation}
where $\eta =$ 2.5, $\delta =$ 45 and $\log_{\rm 10}$P$_{\rm min} =$ 1.  The normalization 
${\int}$f$_{\rm P}$d($\log_{\rm 10}$P) $=$ 1 (where the lower and upper limits of integration are, 
respectively, $\log_{\rm 10}$P$_{\rm min}$ and $\log_{\rm 10}$P$_{\rm max}$, with 
$\log_{\rm 10}$P$_{\rm max} =$ 8.43 and P is in days) is imposed for a binary fraction of 
unity.\footnote{Note that all logarithms in this paper are to the base 10.}  
The assumption of high binary fractions near unity maximizes 
the statistical significance of our analysis of the evolution of the binary populations 
in energy-space.  Additionally, previous work \citep{marks12,marks14} showed that 
such a high binary fraction is needed to obtain the required agreement with the observed 
binary fractions in young Galactic open clusters, when combined with our choice of orbital 
parameter distributions \citep{kroupa95,kroupa95b,kroupa13} and high initial densities (see 
below).  

We run models at Galactocentric radii R$_{\rm GC}$ of 4 kpc, 8 kpc and 10 kpc, 
all with circular orbits in the Galactic potential.  To model the Galactic potential, MOCCA 
assumes a point-mass with total mass equal to the enclosed Galaxy mass at R$_{\rm GC}$.  As 
described in \citep{giersz13}, the criterion used to define the escape rate of stars and 
binaries from the cluster is taken from \citet{fukushige00}, and accounts for the possibility 
that objects with energies greater than the local escape energy can be scattered back into the 
cluster to become re-bound \citep{baumgardt01}.  All models are evolved for 
12 Gyr.\footnote{Models with an initial mass 10$^4$ M$_{\odot}$
do not survive the full 12 Gyr, dissolving after $\sim 10-11$ Gyr of cluster evolution.}  

We adopt an initial mass function (IMF) taken from \citet{kroupa93}
in the mass range 0.08 - 100 M$_{\odot}$.  That is, we assume a mass function of the form 
$\xi$(m) $\propto$ m$^{-\alpha}$, with $\alpha =$ 2.7 for stars with masses m $\ge$ 1 M$_{\odot}$, 
$\alpha =$ 2.2 for stars with masses in the range 0.5 $\le$ m/M$_{\odot}$ $<$ 1, and 
$\alpha =$ 1.3 for stars with masses in the range 0.08 $\le$ m/M$_{\odot}$ $<$ 0.5.  We 
also check that our results are 
roughly\footnote{Our results do slightly change assuming a two-segmented Kroupa IMF 
relative to the three-segmented case, since the former models undergo more stellar
evolution-induced mass loss early on due to having a larger fraction of high-mass stars.
This enhanced mass loss can result in a slightly accelerated cluster evolution, yielding
final binary fractions at 12 Gyr that are slightly lower (by $\sim 1$\%) than we find
adopting a three-segmented IMF.} insensitive to our choice of IMF by re-running all models
with a two-segmented Kroupa IMF taken from \citet{kroupa08}.  We assume a metallicity of 
$Z = 0.001$ for all models.  Initial conditions for all models are 
summarized in Table~\ref{table:initial-MC}. 

For all models, we adopt a King density profile with initial concentration 
W$_{\rm 0} = 6$.  All models are either initially tidally filling or under-filling.  The degree of
under-filling is set by the parameter f$_{\rm und} =$ r$_{\rm t}$/r$_{\rm h}$,
where r$_{\rm t}$ and r$_{\rm h}$ are the tidal and half-mass radii,
respectively.  For tidally-filling models, the parameter f$_{\rm und}$ is defined by
the initial concentration W$_{\rm 0}$.  We consider mass-radius relations of the 
form r$_{\rm h}$ $=$ $\beta$M$_{\rm clus}^{\alpha}$ for some constants $\beta$ and $\alpha$.  For the initially 
tidally-filling models, we adopt $\alpha = 1/3$, corresponding to a constant 
initial density for every set of models with the same initial conditions (but different initial 
total cluster masses).\footnote{Note that 
$\beta \propto$ f$_{\rm und}^{-3}$R$_{\rm GC}^{3}$ if $\alpha = 1/3$, where R$_{\rm GC}$ is the
Galactocentric distance.}  For the initially tidally-underfilling models, we adopt a slightly smaller 
value $\alpha \lesssim 1/3$, corresponding to initial densities that increase slightly (i.e. less 
than a factor of two) with increasing initial cluster mass.  For the tidally-filling 
and tidally-underfilling models, the corresponding initial densities range from 
$\sim 10-100$ M$_{\odot}$ pc$^{-3}$ and $\sim$ 10$^4$-10$^6$ M$_{\odot}$ pc$^{-3}$, 
respectively, inside the half-mass radius.

\begin{table*}
\begin{tabular}{|c|c|c|c|c|c|}
\hline
Total Cluster Mass   &      Time     &  R$_{\rm GC}$  &  f$_{\rm und}$  &     Binary        &   Symbol   \\
  (in M$_{\odot}$)   &    (in Gyr)   &   (in kpc)     &                 &     Fraction      &            \\
\hline
10000          &      10-11       &     10    &   6.8    &  95 &  Open circle     \\
               &                  &           &   50.0   &     &  Filled circle   \\
               &                  &      8    &   6.8    &     &  Open square     \\
               &                  &           &   50.0   &     &  Filled square   \\
50000          &       12         &     10    &   6.8    &  95 &  Open circle     \\
               &                  &           &   50.0   &     &  Filled circle   \\
               &                  &      8    &   6.8    &     &  Open square     \\
               &                  &           &   50.0   &     &  Filled square   \\
100000         &       12         &     10    &   6.8    &  95 &  Open circle     \\
               &                  &           &   50.0   &     &  Filled circle   \\
               &                  &      8    &   6.8    &     &  Open square     \\
               &                  &           &   50.0   &     &  Filled square   \\
500000         &       12         &     10    &   6.8    &  95 &  Open circle     \\
               &                  &           &   50.0   &     &  Filled circle   \\
               &                  &      8    &   6.8    &     &  Open square     \\
               &                  &           &   50.0   &     &  Filled square   \\
               &                  &           &  100.0   &     &  Open pentagon   \\
               &                  &      4    &   6.8    &     &  Open triangle   \\
               &                  &           &   50.0   &     &  Filled triangle \\
               &                  &           &  100.0   &     &  Filled pentagon \\
800000         &       12         &     10    &   6.8    &  95 &  Open circle     \\
               &                  &           &   50.0   &     &  Filled circle   \\
               &                  &      8    &   6.8    &     &  Open square     \\
               &                  &           &   50.0   &     &  Filled square   \\
               &                  &           &  100.0   &     &  Open pentagon   \\
               &                  &      4    &   6.8    &     &  Open triangle   \\
               &                  &           &   50.0   &     &  Filled triangle \\
               &                  &           &  100.0   &     &  Filled pentagon \\
\hline
\end{tabular}
\caption{Initial conditions for all Monte Carlo (MOCCA) models.}
\label{table:initial-MC}
\end{table*}

\subsection{``Observing'' the models} \label{observing}

In order for comparisons to the observed data to be meaningful, the simulated cluster 
properties must be calculated analogously to the observed values from \citet{milone12}.  
That is, we must ``observe'' the simulated clusters in the same way as was done for the 
observations.  To this end, both the core and half-light radii are 
calculated from the 2-D surface brightness profiles of the models.  The
core radius is defined as the distance from the cluster centre at which the
surface brightness falls to half its central value, and the half-light radius 
is defined as the distance from the cluster centre containing half the total 
cluster luminosity.\footnote{We have checked that our results hold qualitatively using instead 
the 3-D (i.e. deprojected) radii and number counts.}  We further ensure that 
the binary fractions are consistently calculated over the range of binary mass 
ratios (q $>$ 0.5) and MS stellar masses (0.47 - 0.76 M$_{\odot}$) used to derive 
the observed values.  

\section{Results} \label{results}

In this section, we present the results of our MOCCA simulations for globular cluster evolution.  We
begin by comparing the simulated binary fractions for \textit{all} models to the observed
binary fractions taken from \citet{milone12}, both inside and outside the half-mass radius.  We
then apply our extrapolation technique from \citet{leigh12} to two sets of models, each with a 
different initial 
mass-radius relation (and hence different initial density) which we refer to as the tidally-filling 
and tidally-underfilling cases.  \textit{Note that we refer to each group of models with the same initial 
binary fraction, orbital parameter distributions and mass-radius relation (defined either by 
whether or not the clusters are initially tidally-filling or tidally-underfilling, and the 
Galactocentric distance), but different initial total cluster masses, as a ``set''.}  

\subsection{Comparisons to the observed binary fractions} \label{binfracs}

\begin{figure*}
\begin{center}
\includegraphics[width=\columnwidth]{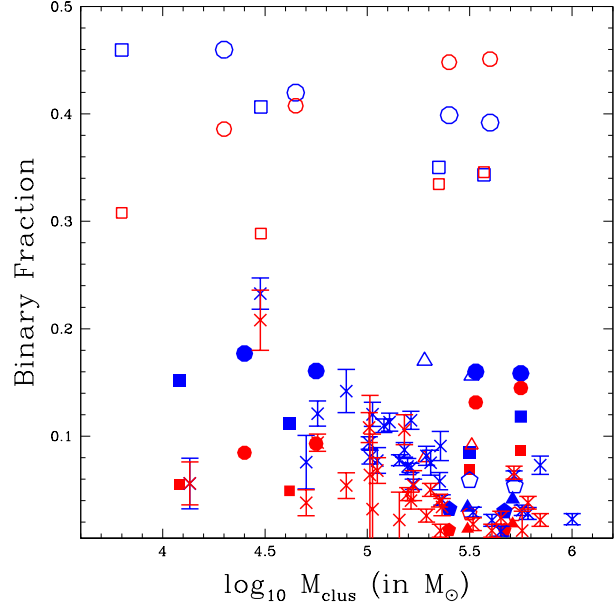}
\end{center}
\caption[Simulated and observed binary fractions inside and outside the half-mass
radius]{The simulated binary fractions are shown for all models at 12 Gyr as a function
of the total cluster mass (in M$_{\odot}$) at 12 Gyr, both inside (blue points) and outside
(red points) the cluster half-mass radius.  Initially tidally-filling and
tidally-underfilling models are indicated by the open and filled symbols, respectively.
Models at 10 kpc, 8 kpc and 4 kpc are shown by the circles, squares and triangles,
respectively.  Models with f$_{\rm und} =$ r$_{\rm t}$/r$_{\rm h} =$ 100 initially 
(i.e. even more tidally-underfilling 
than our standard tidally-underfilling models) are shown by the pentagons for models
evolved both at 8 kpc (open symbols)
and 4 kpc (filled symbols).  The observed binary fractions taken from \citet{milone12} and/or 
calculated in \citet{leigh13b} are 
indicated by the crosses, and error bars show the corresponding uncertainties where available.
\label{fig:fig1}}
\end{figure*}

Figure~\ref{fig:fig1} compares the simulated and observed present-day binary fractions,
both inside (blue points) and outside (red points) the half-mass radius (see the figure
caption for an explanation of which symbols correspond to which models).  
To calculate the observed binary fractions inside r$_{\rm h}$ from
the data provided in \citet{milone12}, we use the relation:
\begin{equation}
\label{eqn:obsbin}
f_{\rm b,h} = \frac{f_{\rm b,c}M_{\rm c}+f_{\rm b,ch}(M_{\rm h}-M_{\rm c})}{M_{\rm h}},
\end{equation}
where f$_{\rm b,c}$ is the core binary fraction, f$_{\rm b,ch}$ is the binary fraction in
the annulus between the core and half-mass radii, M$_{\rm c} =$ 4$\pi$r$_{\rm c}^3$$\rho_{\rm c}$/3
is the mass within the core (assuming a mass-to-light ratio of 2 for all GCs to convert the central 
luminosity densities given in \citet{harris96} to central mass densities) and 
M$_{\rm h} =$ M$_{\rm clus}$/2 is the total mass inside the half-mass radius (and we have assumed 
that the average single star mass $\bar{m}$ is the same everywhere within r$_{\rm h}$, and the average 
binary mass is 2$\bar{m}$).  Importantly, we 
performed the subsequent analysis using all available binary fractions (f$_{\rm b,c}$, f$_{\rm b,ch}$ 
and f$_{\rm b,t}$), instead of just the binary fractions inside and outside r$_{\rm h}$, to verify that 
our key conclusions remain unchanged.
%

As is clear from Figure~\ref{fig:fig1}, the initially tidally-filling models yield binary
fractions at 12 Gyr that are larger than the observed binary fractions by a factor of a few 
(i.e. $\sim$ 2-4) for 
our choice of initial binary fraction (i.e. 95\%).  This is no surprise, because binary disruption 
is too inefficient in such initially extended clusters with such high binary fractions, even with a 
significant soft binary component.  The initially tidally-underfilling models, 
however, yield present-day binary fractions that agree quite well with the observed range,
both inside and outside the half-mass radius.  This illustrates that a high initial cluster
density ($\sim$ 10$^4$-10$^6$ M$_{\odot}$ pc$^{-3}$ at the half-mass radius, relative to
$\sim 10-100$ M$_{\odot}$ pc$^{-3}$ for the tidally-filling models) is needed to
reproduce the observed present-day binary fractions for our choice of orbital parameter 
distributions \textit{if} the initial 
binary fraction was near unity.  This is similar to the results found in \citet{marks11}, however 
the initial densities required by our models are lower than those in \citet{marks11} by 
at least two orders of magnitude (i.e. initial densities on the order of 
$\sim$ 10$^8$ M$_{\odot}$ pc$^{-3}$).  This discrepancy can be accounted for by the inclusion of 
the two-body relaxation phase of cluster evolution in MOCCA, not accounted for by 
\citet{marks11}.

Importantly, this proves neither that the initial density was
indeed high, nor that the initial binary fractions were close to unity.  For example, 
in \citet{leigh13c}, we showed that the present-day central binary fractions can be reproduced 
assuming an universal initial binary fraction of 10\% and a period distribution flat 
in the logarithm of the binary semi-major axis, as suggested by \citet{sollima08}.  Thus, our 
results \textit{for the binary fractions inside r$_{\rm h}$} demonstrate only that the presently 
available data for Milky Way GC binary fractions 
are roughly consistent with an initial binary fraction near unity, in addition
to an universal set of initial binary orbital parameter distributions resembling that
of \citet{kroupa95b} (and in Kroupa et al. 2013), provided the initial
mass-radius relation is chosen to ensure a sufficiently high initial cluster density.  Our 
results for the binary fractions \textit{outside r$_{\rm h}$} tell a different story, however.  
The key feature in Figure~\ref{fig:fig1} we wish to draw the reader's attention to is that the 
\textit{high initial densities adopted in the tidally-underfilling models are needed to reproduce the 
anti-correlation between f$_{\rm b,t}$ and the total cluster mass seen in the observed data}, 
at least for our choice of initial conditions.\footnote{Our results do not necessarily exclude all 
other combinations of initial conditions, only that these initial conditions are able to 
match the observations at 12 Gyr.}  Indeed, for the tidally-filling models, a \textit{correlation} 
is obtained between f$_{\rm b,t}$ and M$_{\rm clus}$ at 12 Gyr.  Thus, even if we were to lower the 
initial binary fractions in the tidally-filling models to ensure better agreement between the 
simulated binary fractions inside r$_{\rm h}$ and the observations, we would still fail to 
reproduce the observed dependence of f$_{\rm b,t}$ on M$_{\rm clus}$ (i.e. outside the half-mass 
radius).  We will return to this important point in Section~\ref{discussion}.

\begin{figure*}
\begin{center}
\includegraphics[width=\columnwidth]{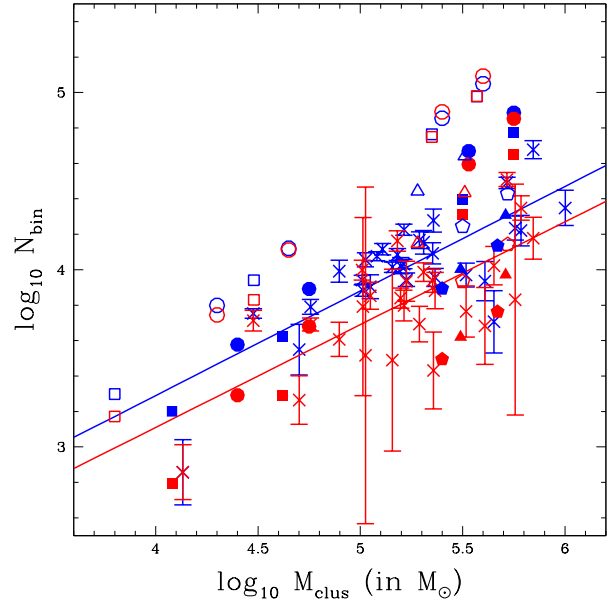}
\end{center}
\caption[Simulated and observed number of binaries inside and outside the half-light radius]{The
(logarithm of) the number of binaries are shown for all models as a function of the total cluster 
mass at 12 Gyr.  
The symbols used to represent each model are the same as in Figure~\ref{fig:fig1}, and
the observed numbers are again indicated by the crosses.  Lines of best-fit are shown
by the solid lines for the observations only, both inside (blue) and outside (red) the
half-light radius.  Error bars are calculated using the binary fraction uncertainties from 
\citet{milone12}.
\label{fig:fig2}}
\end{figure*}

More quantitatively, we obtain lines of best-fit for the observed and simulated relations.
To do this, we first convert the observed and simulated binary fractions to number
counts, both inside and outside the half-mass radius, and then perform least-squares fits
to the data.  The conversion to number counts introduces some additional uncertainty in the
comparison between the models and observations that we are not able to accurately quantify.  
Consequently, we limit ourselves to a qualitative discussion of the uncertainties for the fit 
parameters.  The lines of best-fit are shown in Figure~\ref{fig:fig2} for the 
observations, for which we obtain a sub-linear slope of 0.58 both inside and outside the
half-mass radius (and y-intercepts of 0.92 and 0.79, respectively).  The slopes we
find for the models tend to be much higher, except for the densest models at 4 kpc (see
the filled pentagons in Figure~\ref{fig:fig2}).  This is further evidence that \textit{high initial
densities are needed to reproduce the observations for our choice of initial conditions.}  
Importantly, we note that the correlations are stronger for the models than for the
observed data, with significant scatter about the lines of best-fit.  This is to be expected
since, for example, we adopt the same Galactocentric distance 
for every set of models (i.e. with the same initial conditions but different initial cluster 
masses), and circular orbits.  Real Galactic GCs, however, show a range of 
Galactocentric distances, and often deviate significantly from circular orbits 
\citep[e.g.][]{webb13,webb14}.  
Some of the scatter is also stemming 
from the fact that the dynamical (and absolute) ages of Galactic GCs do not scale 
completely smoothly with total cluster mass (even assuming an universal initial mass-density 
relation) due to, for example, a variable tidal field \citep{webb14}.

At 12 Gyr, the half-mass radii are 
roughly the same as their initial values for the tidally-filling models, whereas
they have increased by a factor of $\sim$ 3-7 in the tidally-underfilling models.
Simultaneously, the total cluster masses have dropped by a factor of a few in
both cases, albeit slightly more so for the tidally-filling models (by a factor of
about two).  It follows that, by 12 Gyr, the mean densities inside r$_{\rm h}$ have
dropped by a factor of $\sim$ a few and a couple orders of magnitude, respectively, for the
tidally-filling and tidally-underfilling models.  Thus, at 12 Gyr, it is the initially 
tidally-underfilling models that best reproduce the observed densities of Galactic GCs.
For example, for our models at 10 kpc, the mean densities inside r$_{\rm h}$ are
an order of magnitude higher in the tidally-filling models (i.e. $\sim$ 100 M$_{\odot}$ pc$^{-3}$)
than in the tidally-underfilling models (i.e. $\sim$ 10 M$_{\odot}$ pc$^{-3}$).  At 4 kpc,
these densities are about an order of magnitude higher at 12 Gyr, however the difference
between the tidally-filling and tidally-underfilling models remains about the same (i.e. an
order of magnitude).  Thus, the present-day mean density inside the half-mass radius offers
another means of constraining the initial cluster density.  Indeed, the observed present-day
mean densities inside r$_{\rm h}$ are typically in the range
$\sim$ 10$^2$-10$^3$ M$_{\odot}$ pc$^{-3}$, which seem to agree better with our initially
tidally-underfilling models.

\begin{figure*}
\begin{center}
\includegraphics[width=\columnwidth]{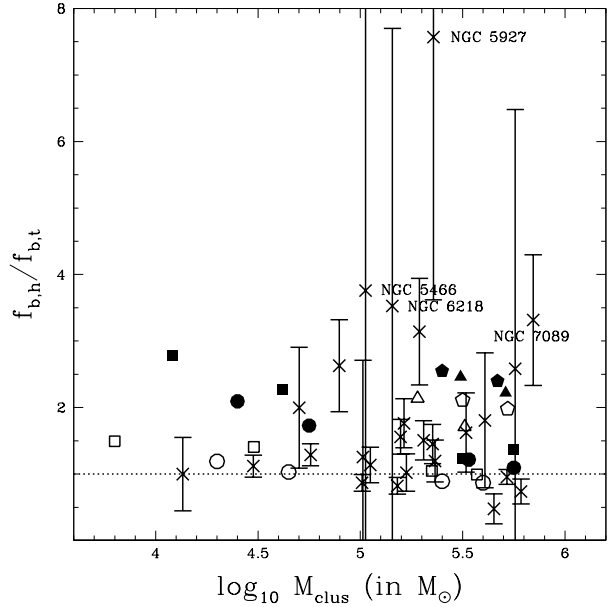}
\end{center}
\caption[Simulated and observed ratios of the binary fractions inside and outside the
half-mass radius]{The ratio f$_{\rm b,h}$/f$_{\rm b,t}$ at 12 Gyr is shown for all
models as a function of the total cluster mass, where f$_{\rm b,h}$ and f$_{\rm b,t}$
denote the binary fractions inside and outside the cluster half-mass radius, respectively.
The symbols used to represent each model are the same as in Figure~\ref{fig:fig1}, and
the observed ratios are again indicated by the crosses.  Error bars are calculated using 
the observed binary fraction uncertainties provided in \citet{milone12}.  The dotted line 
indicates a ratio of unity between the inner and outer binary fractions.
\label{fig:fig3}}
\end{figure*}

But can we simultaneously reproduce the correct binary fractions both inside and outside the
half-mass radius in all clusters?  To address this question, Figure~\ref{fig:fig3} compares the
ratio of the simulated and observed binary fractions inside
and outside the half-mass radius, or f$_{\rm b,h}$/f$_{\rm b,t}$.  The symbols are the
same as in Figure~\ref{fig:fig1}, and we include a (dotted) line to highlight a ratio of unity.
Figure~\ref{fig:fig3} illustrates that our models (both the tidally-filling and tidally-underfilling
models) do indeed roughly reproduce the range of ratios
f$_{\rm b,h}$/f$_{\rm b,t}$ observed in \textit{most} Galactic GCs, with the exception of a
few outliers.  We note a shallow but clear anti-correlation between the total (present-day) cluster 
mass and the ratio f$_{\rm b,h}$/f$_{\rm b,t}$ in the simulations, which is much weaker (but arguably 
still present) in the observations (ignoring the outliers).  This is in rough agreement with the 
results of \citet{fregeau09}, namely that the binary fraction at the half-mass radius remains 
about constant as clusters evolve, while the binary fractions inside and outside the half-mass 
radius slowly increase and decrease, respectively.  Thus, the ratio f$_{\rm b,h}$/f$_{\rm b,t}$ 
should increase with increasing dynamical age.

Some of the more massive Galactic GCs in our sample have
f$_{\rm b,h}$/f$_{\rm b,t} <$ 1, while also being slightly lower than we are able to reproduce
in any of our models.  For example, NGC 6205 and 
NGC 5272 have the lowest values of the ratio f$_{\rm b,h}$/f$_{\rm b,t}$, with values 
of 0.48 and 0.68, respectively.  With the exception of NGC 6205, all of these clusters have 
binary fractions inside the core that exceed their binary fractions in the annulus separating the 
core and half-mass radii (which typically a few parsecs thick and constitutes $\sim$ 40\% of the 
total cluster mass).  Hence, 
the ratio f$_{\rm b,h}$/f$_{\rm b,t}$ is artificially decreased 
in these clusters due to our calculation of f$_{\rm b,h}$ in Equation~\ref{eqn:obsbin} combined 
with the fact that the volume of the core is much smaller than that of the annulus separating the 
core and half-mass radii.  In all but two of these clusters, the binary fraction in the core is 
greater than that outside the half-mass radius.  The exceptions to this are NGC 6101 and NGC 6205.  
These clusters have half-mass relaxation times of a few Gyr \citep{harris96}, which 
could contribute to low dynamical ages for these clusters and hence high binary fractions outside 
r$_{\rm h}$ at 
12 Gyr.  At the same time, these clusters could host a large proportion of soft binaries that are easily 
disrupted in the higher density central cluster regions, which would contribute to low binary 
fractions inside r$_{\rm h}$ \citep{geller13}.  
\footnote{This is especially true if clusters are near a state of core-collapse (either
just before, during and especially after), which can drastically increase the rate of binary
disruption near the cluster centre.  In fact, the number of primordial binaries can become very low, 
such that the number of newly formed binaries begins to outweigh it.}
Given that in these two clusters the ratio f$_{\rm b,h}$/f$_{\rm b,t}$ is only slightly smaller
than seen in our models with the highest initial densities, this discrepancy can partially be 
corrected by adopting slightly higher initial concentrations \citep[e.g.][]{leigh13c}, or if 
these clusters recently underwent a phase of core-collapse (although their present-day concentrations 
do not seem to suggest this).  Having said that, in our models, the exact cluster age at which 
core-collapse first occurs is sensitive to the initial random seed, and core-collapse does not occur 
in any of our simulated clusters.  Therefore, given that the discrepancy is small, our simulations could 
possibly reproduce the observed ratio f$_{\rm b,h}$/f$_{\rm b,t}$ in either NGC 6101 or NGC 6205 
with the right random seed.  This requires verification in future work.  


Interestingly, $\sim 4$ Galactic GCs in our sample show f$_{\rm b,h}$/f$_{\rm b,t}$ values
that are higher than any reproduced by our models.  Basic theory cannot explain this discrepancy 
since extrapolating the results of our 
models to higher initial concentrations should yield lower values for the ratio
f$_{\rm b,h}$/f$_{\rm b,t}$.  We will return to this curious result in Section~\ref{discussion}.  
For now, we note that, when the observational error bars are considered, only one of these 4 
clusters is discrepant with our models, namely NGC 5927.  

\subsection{Quantifying the degree of dynamical processing in energy-space} \label{method}

In this section, we consider how the underlying 
binary orbital parameter distributions are affected by dynamical processing.  In this regard, 
what we are able to learn from the binary fractions alone is limited.  Specifically, binary 
fractions can only tell 
us about the degree of dynamical processing if we know a priori what were the initial 
binary fractions and cluster conditions.  The evolved distributions of binary orbital 
parameters (if available), on the other hand, tell a more detailed story.  

In the subsequent sections, we describe our technique for extrapolating the binary orbital parameter 
distributions obtained at 12 Gyr to any cluster mass using only a handful of models.  The 
method is adapted from \citet{leigh12}, and is meant to facilitate more efficient coverage of the 
relevant parameter space, and ultimately simulate a larger range of initial conditions for 
comparison to the observed data while minimizing the total computational expense.  
We apply this technique to the results of our models in order to better study 
\textit{which} binaries are most affected by dynamical processing, for our choice 
of initial binary orbital parameter distributions.

\subsubsection{Description of the extrapolation technique} \label{extrap}

First, both the rates of two-body relaxation and binary encounters depend 
on the total cluster mass and density.  Specifically, the rate of two-body relaxation increases with 
decreasing cluster mass and increasing density, whereas the rate of binary encounters increases with 
increasing cluster mass and increasing density (ignoring the dependence on the present-day binary 
fraction, which also scales with the cluster mass as, for example, 
f$_{\rm b,core} \propto$ M$_{\rm core}^{-0.4}$) \citep[e.g.][]{sollima08,milone12,leigh13b}.  Moreover, 
the hard-soft boundary is primarily determined 
by the cluster velocity dispersion, which in turn is determined by the total 
cluster mass via the virial theorem (if the cluster is in virial equilibrium, i.e. sometime 
after the gas-embedded phase).  In fact, the rates of two-body relaxation and binary 
encounters should depend \textit{solely} on the total cluster mass \textit{if some initial 
mass-radius and mass-concentration relations are clearly defined} (assuming as well that the 
initial mass function and binary properties are either universal or depend directly on the total 
cluster mass).  It follows from this key assumption that the time evolution of the binary orbital 
parameter distributions also depend directly on the total cluster mass.  Thus, assuming universal 
initial distributions, the present-day binary orbital parameter distributions should vary 
\textit{smoothly} with total cluster mass \textit{for any set of coeval clusters}.  

Next, we describe how we exploit this strong cluster mass-dependence through our extrapolation 
technique.  This 
is feasible for Galactic globular clusters, since they all have roughly the same age (to within 
roughly a couple Gyr) \citep[e.g.][]{salaris02,deangeli05,marin-franch09,hansen13}, but not 
necessarily the same dynamical age.  We emphasize that, for every set of models (i.e. with the same 
initial conditions but different total cluster masses), the initial values for all model parameters 
(e.g. the mass-density relation, the IMF, the Galactocentric distance, etc.) are chosen to ensure 
that the rates of two-body relaxation and binary encounters, and hence the present-day cluster-to-cluster 
differences in the binary orbital parameter distributions, depend solely on the total cluster mass.  
In other words, the dynamical ages for every set of models sharing the same initial conditions depend 
only on the total cluster mass.

We begin by converting the present-day binary orbital 
parameter distributions into orbital energy distributions.  This is because, as described in 
Section~\ref{intro}, binary 
encounters most directly affect the binary orbital parameter distributions in energy 
space, whereas two-body relaxation does not care about the binary orbital parameters (only the 
total binary mass).  
Then, we bin the present-day orbital energy distributions such that the bin sizes are 
equal in log-space, each spanning 1 dex.  This is done for every model, and then models 
with the same initial conditions are grouped together.  Hence, each group contains 
5 models, each with a different initial total cluster mass but otherwise identical initial 
conditions.  

Now, in order to quantify 
cluster-to-cluster differences in the present-day orbital energy distributions, we normalize 
the distributions by dividing each orbital energy by the average single star kinetic energy in the 
cluster at 12 Gyr, or E$_{\rm avg} =$ 0.5m$\sigma^2$, where m is the average single star mass (taken 
to be 0.35 M$_{\odot}$ for all models, since the average mass is within the range 0.3 - 0.4 M$_{\odot}$ 
at 12 Gyr in all models) and $\sigma$ is the root-mean-square 
velocity \citep{binney87}:
\begin{equation}
\label{eqn:sigma}
\sigma = \Big( \frac{2GM}{5r_{\rm h}} \Big)^{1/2},
\end{equation}  
where M is the total cluster mass and r$_{\rm h}$ is the half-mass radius.  Thus, in log-space, 
a \textit{normalized} value for the orbital energy of 0 corresponds to the 
hard-soft boundary in each cluster.  Next, we obtain 
for each group of models (with the same initial conditions but different initial total cluster 
masses) lines of best-fit for 
(the logarithm of) the number of binaries belonging to each orbital energy bin
versus (the logarithm of) the total number of binaries spanning \textit{all} energy 
bins, which corresponds to the total number of binaries 
in a given cluster.\footnote{We also repeat this procedure using only those binaries that 
satisfy the observational criteria of \citet{milone12}, however our results are 
qualitatively the same.}  This can be written:
\begin{equation}
\label{eqn:frac-bin}
\log_{10} N_{\rm en,i} = {\gamma_i}\log_{10} N_{\rm bin} + \delta_i,
\end{equation}
where N$_{\rm en,i}$ is the number of binaries belonging to orbital energy bin $i$, 
N$_{\rm bin}$ is the total number of binaries in the cluster at 12 Gyr (i.e. spanning all orbital 
energy bins), and $\gamma_{\rm i}$ and $\delta_{\rm i}$ are both constants.  

If the fraction of binaries belonging 
to each orbital energy bin, or f$_{\rm en,i}$ $=$ N$_{\rm en,i}$/N$_{\rm bin}$,
is constant for all cluster masses, then we would expect N$_{\rm en,i}$ to
scale linearly with N$_{\rm bin}$.  Or, equivalently, $\gamma_{\rm i}$ $\sim$ 1 in
Equation~\ref{eqn:frac-bin}.  
However, if there is any systematic dependence of f$_{\rm en,i}$ on the
total cluster mass, then we should find that N$_{\rm en,i}$ does
\textit{not} scale linearly with N$_{\rm bin}$.  In log-log space, the
slope of the line of best-fit for orbital energy bin $i$
should be less than unity (i.e. $\gamma_{\rm i}$ $< 1$) if f$_{ \rm en,i}$
systematically decreases with increasing cluster mass.  Conversely,
we expect $\gamma_{\rm i}$ $> 1$ if f$_{\rm en,i}$ systematically increases
with increasing cluster mass.  \textit{Therefore, the exact values of $\gamma_{\rm i}$ and
$\delta_{\rm i}$ are sensitive to the initial cluster conditions, in particular
the initial binary orbital parameter distributions as well as the initial cluster 
density (i.e. the assumed mass-radius relation).}

Our motivation for adopting this technique is as follows.  Equation~\ref{eqn:frac-bin} quantifies, 
for any cluster mass, the number of binaries belonging to each orbital energy bin as a function of the 
total number of binaries.  The exact values of the parameters in Equation~\ref{eqn:frac-bin} are 
sensitive to the adopted initial binary orbital parameters and initial mass-density relation.  
Equation~\ref{eqn:frac-bin} can 
be converted into an equation that provides the number of binaries in each orbital 
energy bin as a function of the total number of \textit{stars} using the total 
binary fraction, or N$_{\rm bin} =$ f$_{\rm b}$N, where f$_{\rm b}$ is the number of objects 
that are binaries and N is the total number of objects in the cluster (i.e. single stars and 
binaries).  In turn, this can be converted into an expression that depends on the total 
cluster mass M$_{\rm clus}$ using the simple relation M$_{\rm clus}$ $=$ mN.  Thus, 
Equation~\ref{eqn:frac-bin} effectively 
provides a means of quantifying cluster-to-cluster differences in the distribution of binary orbital 
energies as a function of the total (present-day) cluster mass.  It follows from this that, 
within the universality hypothesis (i.e. adopting universal initial binary properties and an 
universal initial mass-density relation), knowing only the present-day observed total cluster 
mass and binary fraction, one can infer the underlying distributions of orbital energies for some 
assumed set of initial conditions.  This 
is qualitatively similar to the technique presented in \citet{marks11} for young star clusters 
at ages between 0-5 Myr.  \textit{Although we only apply the technique presented in this paper 
to two particular sets of initial conditions, in principle it can be done for any set of 
universal initial binary properties and mass-density relations.}  



\subsubsection{Application of the extrapolation technique} \label{app_extrap}

In this section, we apply our extrapolation technique to two particular sets of models as a 
proof-of-concept, each with a different initial mass-density relation but otherwise 
identical initial conditions.  The densest models are initially tidally-underfilling 
whereas the sparsest models are initially tidally-filling, and all models are evolved 
at a Galactocentric distance of 10 kpc.  We remind the reader that we adopt an 
initial binary fraction of 95\% for all models.  This serves to maximize the final 
binary number counts at 12 Gyr.  This is particularly important in those models with 
the highest initial densities (i.e. the tidally-underfilling models), since a significant 
number of soft binaries are disrupted 
very early on in the cluster lifetime (before the clusters have had enough time to expand 
to fill their tidal radii), causing the number of single stars to rapidly increase along 
with a corresponding decrease in the binary fraction.  Additionally, we consider \textit{all} 
binaries in this section, as opposed to just those binaries along the MS.

\begin{figure*}
\begin{center}
\includegraphics[width=\columnwidth]{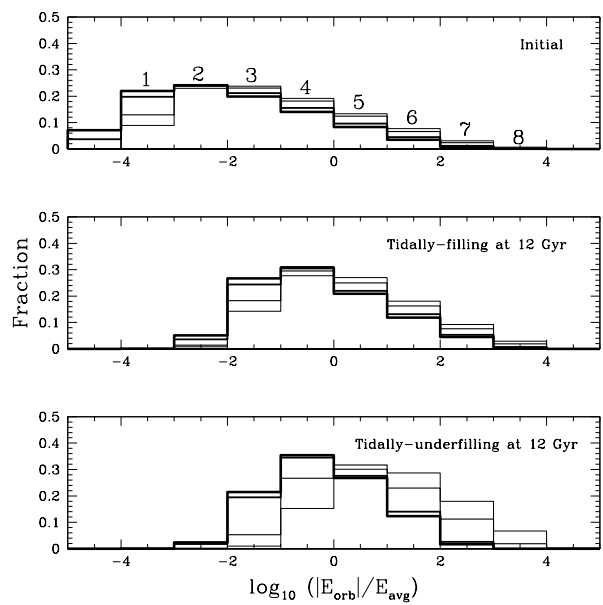}
\end{center}
\caption[Initial and final binary orbital energy distributions, both for the 
tidally-filling and tidally-underfilling models at 12 Gyr]{The top panel shows 
the initial (i.e. t $=$ 0) distribution of binary orbital energies for clusters 
with total initial masses of 5 $\times$ 10$^4$, 10$^5$, 
5 $\times$ 10$^5$ and 8 $\times$ 10$^5$ M$_{\odot}$.  Note that, if the orbital energies 
were not normalized by the average kinetic energy in each cluster, the distributions would 
all be identical at t $=$ 0.  Due to our normalization, more massive models have a larger 
fraction of soft binaries due to the dependence of the hard-soft boundary on the cluster 
mass.  The middle and bottom 
panels show the orbital energy distributions at 12 Gyr for the tidally-filling 
and tidally-underfilling models, respectively.  The width of the lines forming 
each histogram correlate with the initial total cluster masses, such that the 
thickest line had the largest masses.  The orbital energies 
have been normalized by the average single star kinetic energy or 
E$_{\rm avg} =$ 0.5m$\sigma^2$, where m is the average single star mass (taken 
to be 0.35 M$_{\odot}$ for all models) and $\sigma$ is the root-mean-square 
velocity.  Thus, 0 on the x-axis corresponds to the hard-soft boundary 
in each cluster.  Each histogram bin has been normalized by the total number of 
binaries (at 12 Gyr) in the corresponding cluster.  The numbers shown in the top 
panel indicate the bins in orbital energy used to 
generate Figure~\ref{fig:fig5} and shown in Table~\ref{table:lines}.  
\label{fig:fig4}}
\end{figure*}

Figure~\ref{fig:fig4} shows normalized histograms of the binary orbital 
energy distributions both initially (top panel) and at 12 Gyr, both for the tidally-filling 
(middle panel) and tidally-underfilling (bottom panel) models.  
As is clear from a comparison of the three panels in 
Figure~\ref{fig:fig4}, dynamical processing 
acts to reduce the number of soft binaries, increasing the fraction of hard binaries in the cluster.  
This effect is the most dramatic for the tidally-underfilling case, since the initial densities 
were the highest.  Importantly, this result is roughly insensitive to the initial binary fraction 
due to our normalization technique (which removes the dependence on absolute numbers).  Additionally, 
for every set of initial conditions, it is the 
most massive models that end up the most depleted of binaries, as expected.  

\begin{figure*}
\begin{center}
\includegraphics[width=\columnwidth]{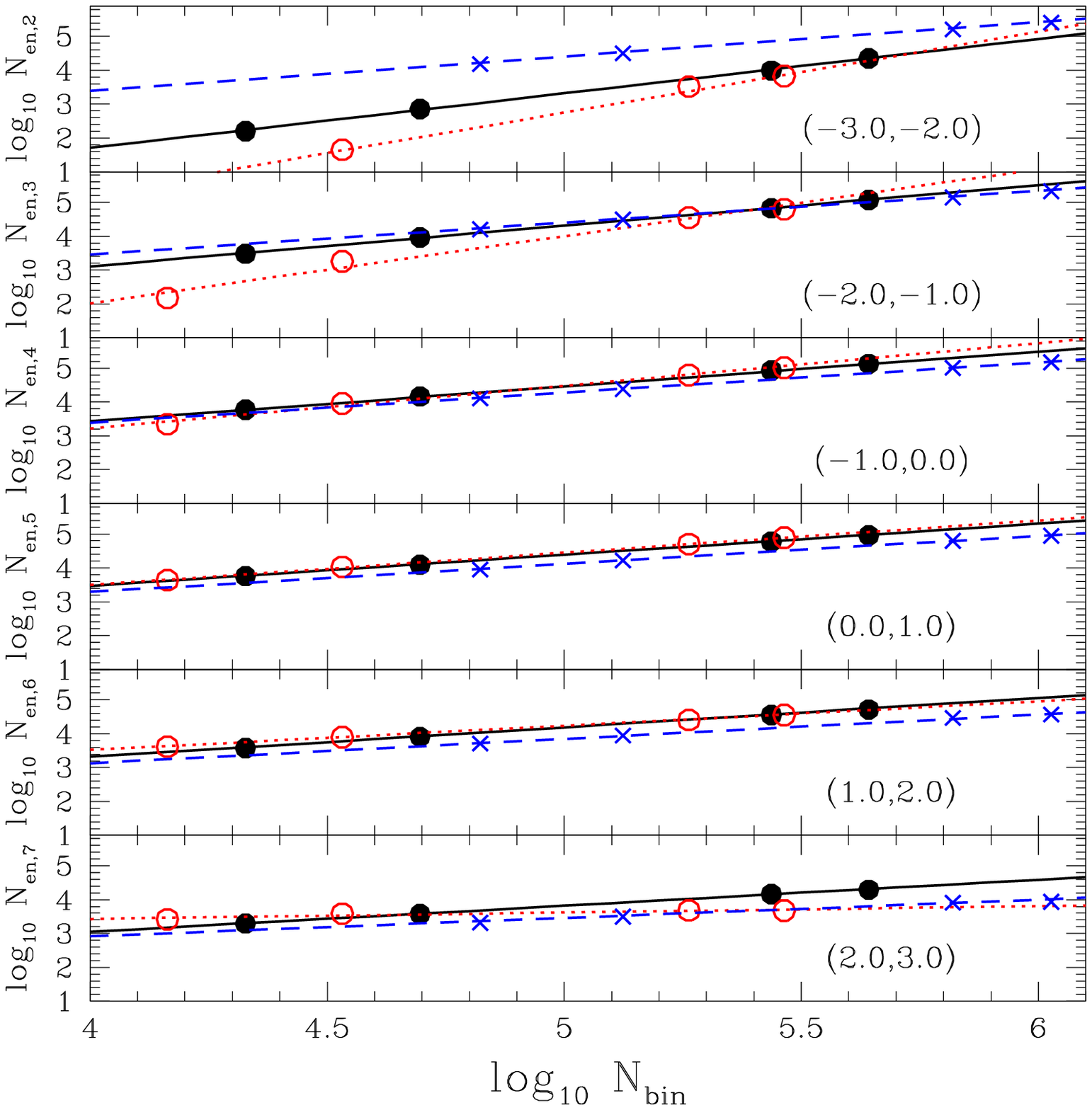}
\end{center}
\caption[Lines of best-fit from our extrapolation technique for the initial (i.e. t $=$ 0) 
distributions as well as those at 12 Gyr in both the tidally-filling and tidally-underfilling 
models]{Each panel shows the number of binaries N$_{\rm en,i}$ 
in the corresponding orbital energy bin (indicated in Figure~\ref{fig:fig4}) 
as a function of the total number of binaries N$_{\rm bin}$ in the cluster, 
for models with total initial masses of 5 $\times$ 10$^4$, 10$^5$, 5 $\times$ 10$^5$ 
and 8 $\times$ 10$^5$ M$_{\odot}$ (we do not include models with 10$^4$ stars initially, 
since they do not survive to an age of 12 Gyr).  The results are shown for the initial 
distributions at t $=$ 0 (blue crosses), as well as at 12 Gyr for both the tidally-filling 
(filled circles) and tidally-underfilling (open red circles) models.  The range in 
(normalized) orbital energies corresponding to bin $i$ are indicated in Figure~\ref{fig:fig4} 
for all six bins shown here, as well as in the lower right of each panel.  Lines of best-fit are 
obtained using Equation~\ref{eqn:frac-bin} 
and shown for each bin in orbital energy (see Table~\ref{table:lines} for the exact fits).  The 
solid black, dotted red and dashed blue lines 
correspond to the initial, tidally-filling and tidally-underfilling cases, respectively.
\label{fig:fig5}}
\end{figure*}

Lines of best-fit are obtained using Equation~\ref{eqn:frac-bin} for each bin in orbital 
energy and provided in Table~\ref{table:lines}.  These lines are 
shown in Figures~\ref{fig:fig5} for the initial distributions (i.e. t $=$ 0) as well as 
for both the tidally-filling and tidally-underfilling models at 12 Gyr.  

\begin{table*}
\begin{tabular}{|c|c|c|c|c|}
\hline
Bin   &    Range   &           Initial       &      Filling            &     Underfilling           \\
      &   ($\log_{10}$ (E$_{\rm orb}$/E$_{\rm avg}$))        &       ($\gamma_{\rm i}$; $\delta_{\rm i}$)    &      ($\gamma_{\rm i}$; $\delta_{\rm i}$)      &       ($\gamma_{\rm i}$; $\delta_{\rm i}$)     \\
\hline
1  &  -4.0;-3.0  &  0.76; 1.91   &   --; --      &   --; --      \\
2  &  -3.0;-2.0  &  1.02; -0.69  &  1.61; -4.73  &  2.39; -9.20  \\
3  &  -2.0;-1.0  &  0.94; -0.30  &  1.20; -1.69  &  1.98; -5.90  \\
4  &  -1.0;0.0   &  0.89; -0.17  &  1.03; -0.69  &  1.26; -1.82  \\
5  &  0.0;1.0    &  0.83; -0.03  &  0.92; -0.21  &  0.95; -0.30  \\
6  &  1.0;2.0    &  0.72; 0.25   &  0.87; -0.16  &  0.72; 0.64   \\
7  &  2.0;3.0    &  0.54; 0.76   &  0.77; -0.03  &  0.19; 2.67   \\ 
8  &  3.0;4.0    &  0.20; 1.74   &   --; --      &   --; --      \\
\hline
\end{tabular}
\caption{Parameters for the lines of best-fit (see Equation~\ref{eqn:frac-bin}) for the initial,
tidally-filling and tidally-underfilling models.  Lines of best-fit are of the form 
$\log_{10}$ N$_{\rm en,i} =$ $\gamma_{\rm i}$$\log_{10}$N$_{\rm bin}$ + $\delta_{\rm i}$, where $\gamma_{\rm i}$ 
and $\delta_{\rm i}$ are both constants.}
\label{table:lines}
\end{table*}

%

One of the key features characteristic of Figures~\ref{fig:fig4} and~\ref{fig:fig5} is that 
\textit{the fraction of hard binaries increases with decreasing total cluster mass,} and this 
effect becomes increasingly significant the more dynamical processing a binary population 
undergoes.  This is illustrated in Figure~\ref{fig:fig4} through our normalization method, since 
we see a clear trend for orbital energy bins with $\log_{10}$ (E$_{\rm orb}$/E$_{\rm avg}$) $>$ 0 
(i.e. hard binaries) to 
become increasingly populated as more dynamical processing occurs.  Similarly, a higher initial 
density results in a 
larger range of slopes and y-intercepts for our lines of best-fit over the six bins in 
orbital energy shown in Figure~\ref{fig:fig5}.  This is evident from Table~\ref{table:lines} upon 
comparing the range of slopes 
and y-intercepts for the tidally-filling (1.61 to 0.77 and -4.73 to -0.03 for the slopes 
and y-intercepts, respectively) and tidally-underfilling (2.39 to 0.19 and -9.20 to 2.67) models.  
This larger range of slopes and y-intercepts seen in initially denser clusters (i.e. the 
tidally-underfilling models) is due to a corresponding increase in the degree of dynamical 
processing, which in turn contributes to the fraction of hard binaries at 12 Gyr increasing 
more steeply with decreasing total cluster mass.  This is the direct result of our assumed initial 
mass-density relation.  
Specifically, we assume a mass-density relation 
corresponding to a constant initial density for every set of models (i.e. with the same
initial conditions).  It follows from this that the encounter rate scales as
$\Gamma \propto$ M$_{\rm clus}^{-1}$.  Including the dependence of the encounter rate on
the binary fraction only steepens the inverse mass-dependence further, since
$\Gamma$ $\propto$ f$_{\rm b}$ and the binary fraction is anti-correlated with the
total cluster mass \citep{sollima08,milone12}.  Thus, the rate of dynamical processing
is fastest in the least massive clusters for every set of models sharing the same
initial conditions.  Consequently, the rate of soft binary disruption is also the
highest in the lowest mass clusters, causing the fraction of hard binaries to increase
with decreasing cluster mass.  Said another way, for a given orbital energy bin, the mean 
separation is larger in less massive 
clusters, which contributes to a shorter binary encounter time.  This is because our normalization
relies on dividing by the orbital energy corresponding to
the hard-soft boundary, which is at larger orbital energies (i.e. less negative) in less massive
clusters.  This effect is the most significant in the initially
tidally-underfilling models, since they have the highest initial densities and hence
the highest encounter rates. 

To summarize, without the need to do additional simulations to increase our range of simulated cluster 
masses, we now know the number of binaries in each orbital energy bin for 
any cluster that shares the same initial conditions (but a different initial total cluster mass) of 
either our tidally-filling or tidally-underfilling models.  In the future, for a given set of initial 
conditions, one needs only simulate clusters over a small range in initial mass to extrapolate the 
results over the entire initial mass range.  Finding the extrapolation of the initial conditions that 
best intersect with actual Galactic GCs in the N$_{\rm en,i}$-N$_{\rm bin}$ phase space will help 
to constrain the true initial cluster conditions, as discussed in Section~\ref{initial}.

\section{Discussion} \label{discussion}

In this section, we discuss the implications of our results for the properties of 
primordial binaries in Galactic globular clusters, within the framework of the universality 
hypothesis.

\subsection{What were the initial binary properties in Galactic globular clusters?} \label{initial} 

Our results illustrate that the observed present-day binary fractions in Galactic 
GCs can, to first-order, be reasonably well reproduced assuming universal initial binary 
properties, specifically a binary fraction near unity, binary orbital parameter 
distributions resembling those of \citet{kroupa95} and an initial mass-radius 
relation with high initial densities (10$^4$ - 10$^6$ M$_{\odot}$ pc$^{-3}$).  In \citet{leigh13c}, we 
demonstrated a similar consistency between the observations and the binary universality 
hypothesis, but adopting a binary fraction of 10\% and an orbital period distribution 
flat in the logarithm of the binary semi-major axis, while exploring a wide range of initial cluster 
densities (10$^2$-10$^6$ M$_{\odot}$ pc$^{-3}$).  Importantly, this was for the \textit{central} binary 
fractions.  Thus, as far as the binary \textit{fractions} inside r$_{\rm h}$ are concerned, the problem of 
whether or not the primordial binary properties 
in GCs were universal is degenerate -- that is, to first-order, the present-day 
observations can 
be reproduced within the universality hypothesis adopting different combinations 
of the initial binary fraction, binary orbital parameter distributions and 
mass-radius relations.  This was also indirectly demonstrated in \citet{hurley07} using $N$-body 
models, who showed from a suite of simulations of up to 10$^5$ stars that the overall binary 
fraction remains close to its primordial value (starting with binary fractions on the order 
of 5\%), but increases markedly in the cluster core.  Importantly, however, these authors included 
(mainly) hard binaries only.  Thus, in these models, binary 
destruction in the core is outweighed by mass segregation delivering binaries to the core 
at a faster rate (combined with binary creation during dynamical interactions).  

So how then can we distinguish between degenerate sets of initial conditions?  As is 
illustrated in Figure~\ref{fig:fig1}, \textit{the dependence of the present-day binary fraction 
f$_{\rm b,t}$ outside the half-mass radius r$_{\rm h}$ on the total cluster mass can break 
degeneracies that arise using the binary fractions inside r$_{\rm h}$ alone.}  
Specifically, assuming an universal initial binary fraction outside r$_{\rm h}$ along 
with universal initial orbital parameter distributions, initially tidally-underfilling clusters 
with high initial densities on the 
order of $\sim$ 10$^4$-10$^6$ M$_{\odot}$ pc$^{-3}$ are needed to reproduce the observed 
anti-correlation between the total cluster masses and the binary fractions in the cluster 
outskirts.  Indeed, lower initial densities yield a \textit{correlation} between f$_{\rm b,t}$ 
and M$_{\rm clus}$, as illustrated in Figure~\ref{fig:fig1} by the initially tidally-filling 
models.  This is in general consistent with the results of \citet{hurley07} and \citet{fregeau09}, 
since the mass 
segregation process operates faster in less massive clusters.  In other words, if 
the overall global binary fraction remains the same while the binary fraction inside
r$_{\rm h}$ increases then the binary fraction outside r$_{\rm h}$ must decrease 
\citep[e.g.][]{fregeau09}, and this process operates the fastest in clusters with the 
shortest half-mass relaxation times.  This 
leads to a \textit{correlation} between the cluster dynamical age (for which we use the total
cluster mass as a proxy in this paper) and the binary fraction outside r$_{\rm h}$.

Our results are also indicative of an initial mass-radius relation that is less steep than 
we have adopted here.  That is, instead of assuming a constant initial density, the 
agreement between our models and the observations would improve if we assumed a mean 
initial density that increases with increasing cluster mass.  
This is required to obtain a better agreement between our models and the observations 
for every individual set of models with the same initial conditions but different 
initial total cluster masses.  As seen in Figure~\ref{fig:fig2}, the observed dependence of 
the number of binaries on the total cluster mass is more sub-linear than we obtain in 
any of our models.  Adopting instead an initial-mass radius relation that corresponds to even 
higher densities in more massive clusters than in our tidally-underfilling models would correct 
this discrepancy.  Thus, assuming 
an initial mass-radius relation of the form r$_{\rm h}$ $=$ $\beta$M$_{\rm clus}^{\alpha}$, our 
results suggest $\alpha <$ 1/3.  This is similar to Equation 7 in \citet{marks12},
who find $\alpha \sim$ 0.13 $\pm$ 0.04.  

To summarize, assuming global initial binary fractions near unity and the initial binary orbital 
parameter distributions of \citet{kroupa95} (i.e. most binaries 
are initially soft), our results are the most consistent with high 
initial cluster densities of the order 10$^4$-10$^6$ M$_{\odot}$ pc$^{-3}$.  We require the highest 
initial densities ($\sim$ 10$^6$ M$_{\odot}$ pc$^{-3}$) to reproduce the observed binary fractions 
in the most massive Galactic GCs in our sample, and a range at slightly lower initial densities 
($\sim$ 10$^4$-10$^5$ M$_{\odot}$ pc$^{-3}$) to match the observations in the lowest mass GCs in 
our sample.  This suggests an initial mass-density relation corresponding to higher initial densities 
in more massive clusters (see above).  These initial 
conditions are able to reproduce the observed present-day distribution of binary fractions 
\textit{both inside and outside the half-mass radius}.  Conversely, adopting initial binary fractions 
near $\sim$ 10\% with a dominant hard binary component combined with comparably high initial densities 
can reproduce the observed binary fractions \textit{inside r$_{\rm h}$ only}.  We 
emphasize that our results do not necessarily exclude other possible combinations of the initial cluster
conditions that might also reproduce the observed data for the Galactic GC population within the
framework of the universality hypothesis, nor have we ruled out non-universal initial conditions.  For 
example, our results suggest that the observed binary fractions outside the half-mass radius 
might be reproducable assuming an universal (global) initial binary fraction of 10\% with a 
dominant hard binary component \textit{if} primordial binary mass segregation is also included in 
some clusters.  With that said, we emphasize that our results, in particular initial cluster 
densities in the range 10$^4$-10$^6$ M$_{\odot}$ are observed in many massive young star-forming 
regions, including the ONC, DR21, M17, RCW36, RCW38, NGC1893, etc. (see, for example, Kuhn et al. 2014).

\subsection{Evolution in energy-space} \label{energy-space}

In this section, we discuss the evolution of the initial binary populations in 
energy-space as a function of the initial cluster density, both for the initially 
tidally-filling and tidally-underfilling cases.  As we have shown, in those cases where 
the binary fractions alone are not sufficient to fully remove all degeneracies in 
the initial conditions (for example, as we have shown, this can be the case inside the half-mass 
radius), this can be done by also comparing the relative fractions of hard 
and soft binaries at 12 Gyr.

First, at 12 Gyr, the fraction of hard binaries has increased relative to the 
initial fraction in all models, as seen in Figure~\ref{fig:fig5}.  This is the case for both the 
tidally-filling and tidally-underfilling models, but the effect is more significant 
for the latter.  This is because a higher initial density translates into a larger degree of
dynamical processing and hence a larger present-day fraction of hard binaries.  Thus, the 
fraction of hard binaries offers one probe of the initial cluster density.  Indeed, 
as already illustrated via our results for the binary fractions, the fraction of 
hard binaries in the cluster outskirts (i.e. beyond the half-mass radius) can offer an 
even more sensitive probe of the initial cluster density.  
Second, the fraction of hard binaries increases with increasing cluster mass at 12 Gyr, for 
every set of models sharing the same initial conditions (but different total cluster masses).  
This is the case for both the tidally-filling and tidally-underfilling models, however the 
effect is once again more significant for the latter.  This is the direct result of 
our assumed initial mass-density relation.  Therefore, the dependence of the hard and soft 
binary fractions on the present-day total cluster mass (and density) can help to constrain 
the initial cluster mass-density relation.

It follows from our example application that our extrapolation 
technique can be used to not only efficiently sample the relevant parameter space of initial conditions,
but also to help break any degeneracies (within the universality hypothesis) between the presently 
observed binary properties and the initial cluster conditions.  For example, consider a 
particular Galactic GC with measured values for both f$_{\rm b,h}$ and f$_{\rm b,t}$.  We
evolve to 12 Gyr different sets of models having the same initial binary fraction, orbital parameter
distributions and mass-density relation (for example, an universal initial binary fraction of 50\%,
an initial density of 10$^3$ M$_{\odot}$ pc$^{-3}$ and some suitable combination of binary orbital
parameter distributions), but different initial total cluster masses.  We are not concerned
with reproducing the exact present-day total cluster mass in our models at 12 Gyr, but instead
generate a range of total cluster masses at 12 Gyr for every set of initial conditions, and then use
our extrapolation technique to calculate the distributions of orbital energies at 12 Gyr for the
precise total mass corresponding to our cluster of interest.  If more than one
set of models yields binary fractions that are consistent with the observed binary fractions \textit{both}
inside and outside the half-mass radius, this degeneracy can be broken using the relative fractions
of hard and soft binaries.  That is, the relative fractions of hard and soft binaries, which
probe the degree of dynamical processing, should differ.  This is quantified by our extrapolation
technique.  Thus, observational constraints for the relative fractions of hard and soft binaries, if
available,\footnote{Or perhaps using some proxies for the numbers of hard and soft binaries.  For example,
the numbers of LMXBs or MSPs can be used as a proxy for the number of hard binaries, and the number of
photometric binaries along the MS can be used as a proxy for the total (i.e. hard and soft) number of
binaries.} can be used to further constrain the initial cluster conditions in those situations where
the present-day binary fractions alone cannot fully do the job.  Depending on the quality of the data and 
the degree of degeneracy among the initial conditions, more (or less) binning may 
be needed to reliably isolate the set of initial conditions that best match the observed data.   
We note as well that, for a more
direct comparison to observed data, our method can just as easily be applied directly to the underlying
period, mass ratio and eccentricity distributions instead of the orbital energy distributions.  It can
also be applied to different spatial subsets of the cluster or different radial bins (e.g. inside and
outside the half-mass radius).

Interestingly, in the initially tidally-underfilling case, the fit is poor for our highest bin in
orbital energy (i.e. bin 8, which is not shown in Figure~\ref{fig:fig5} or Table~\ref{table:lines} since
the simple least-squares fits performed for the other orbital energy bins is not appropriate here).
This seems to be connected to the earliest stages of cluster evolution, before the clusters have had
a chance to expand to fill their tidal radii, since the clusters are initially binary-dominated
such that most soft binaries are destroyed during
the first few crossing times of the cluster lifetime \citep{marks11}.  The models are
initially in virial equilibrium, so that they begin to expand on a two-body relaxation
timescale (ignoring expansion due to stellar evolution-driven mass loss).  It is the most
massive clusters that take the longest to expand to fill their
tidal radii, since the timescale for two-body relaxation increases with cluster mass.  Thus, we
postulate that our models deviate from an universal mass-radius relation during the brief
period after t $=$ 0 but before all clusters fill their tidal radii.  That is, clusters that
are not yet tidally-filling retain their binaries in the outskirts so that they continue
to be affected by dynamical interactions, whereas tidally-filling clusters lose stars and
binaries from their outskirts to the tidal field of the Galaxy.  It is this brief
high-density state that results in the poor fit seen in the lower panel of Figure~\ref{fig:fig5}
for the tidally-underfilling models,
since more of the softest binaries are disrupted in the most massive clusters relative to what is
seen in Figure~\ref{fig:fig5} for the initially tidally-filling models.  This initial phase of
differential expansion is avoided altogether in the initially tidally-filling models.  We 
conclude from this that very soft binaries potentially offer the most sensitive probe of the 
initial cluster conditions.  

Importantly, \textit{this phase of differential expansion seen in the tidally-underfilling models 
contributes to the anti-correlation we see between the binary fractions outside r$_{\rm h}$ and the 
total cluster 
mass.}  In the initially tidally-filling models, we find a correlation between f$_{\rm b,t}$ 
and M$_{\rm clus}$, which is consistent with the results of \citet{hurley07} and \citet{fregeau09}, 
but inconsistent with the observations \citep{milone12}.


%
%

\subsection{Clusters of interest} \label{clusters}

We emphasize that, while the observed binary fractions for most of the clusters in our 
sample can be well-reproduced assuming universal initial binary properties, not 
every cluster fits the mold of binary universality, at least at first glance.  In particular, 
Figure~\ref{fig:fig3} 
suggests that some clusters may be born with initial binary mass segregation, namely 
NGC 5466, NGC 5927, NGC 6218 and NGC 7089.  This is needed to reproduce the high ratios observed 
between the binary fractions inside and outside the half-mass radius.  NGC 5466 has the lowest 
mean density inside r$_{\rm h}$ of all the clusters in our sample, and the difference is greater 
than an order of magnitude ($\sim$ 20 M$_{\odot}$ pc$^{-3}$ compared to several 
100-1000 M$_{\odot}$ pc$^{-3}$).  This is likely due mainly to the relatively large 
Galactocentric distance of this cluster (16.3 kpc), which has allowed it to expand 
considerably over its lifetime \citep[e.g.][]{webb14}.  However, it does not help to explain the 
large ratio 
f$_{\rm b,h}$/f$_{\rm b,t}$ -- unless, perhaps, single stars expand more in the Galactic potential 
than do binaries, contributing to the appearance of primordial binary mass segregation.  NGC 5927 
and NGC 6218 have observed Galactocentric distances of 
4.6 kpc and 4.5 kpc, respectively.  At such small distances from the Galactic centre, 
mass-loss due to tidal effects from the Galaxy begin to play a very important role.  Based 
on a comparison between the results of MOCCA and NBODY6 at a Galactocentric radius of 4 kpc, 
MOCCA appears to be under-estimating the rate of mass-loss across the tidal boundary 
(Madrid, Hurley \& Leigh 2014, in preparation).  Tides are known to most dramatically 
affect the cluster outskirts \citep[e.g.][]{webb14} and, if single stars escape at a 
greater rate than do binaries, this could contribute to lowering the binary fraction outside 
r$_{\rm h}$ relative to the value inside r$_{\rm h}$.  More work is needed to properly 
treat the Galactic potential at small Galactocentric distances, both in MOCCA and in $N$-body 
models, and hence to address this interesting question.  

Another option is that these clusters were born with significant substructure, or in
lower mass clusters that eventually merged.  In this scenario, the lower initial cluster
masses of the sub-clumps translate into shorter two-body relaxation times, and hence shorter 
mass segregation times for binaries relative to single stars.  This could explain the significant
concentration of binaries within r$_{\rm h}$ in these clusters, relative to what we would
expect beginning with a single monolothic collapse scenario.

With that said, when the uncertainties on the observed binary fractions are taken into account 
\citep{milone12}, the number of Galactic GCs with binary fractions we are unable to reproduce 
with our models is very small (with NGC 5927 being the only outlier).  It is worth 
considering whether or not non-members may be contaminating the determination of the binary fractions 
in these few clusters.  Alternatively, multiple stellar populations could be confusing the 
determination of the binary fractions inside and outside r$_{\rm h}$, especially if one of these 
populations has a higher binary fraction or is more centrally concentrated.

\subsection{Limitations of the Monte Carlo approach} \label{accuracy}

In this section, we briefly highlight several key issues inherent to the Monte Carlo 
method for simulating globular cluster evolution.  Although we do not expect any of 
these issues to have significantly affected our results, we list them here for 
completeness.  


For our purposes, the most significant issue is MOCCA's treatment of the Galactic tidal 
field, which is relevant to the rate of escape of single and binary stars across the tidal
boundary, and hence the binary fraction in the cluster outskirts.  Specifically, MOCCA 
approximates the Galaxy as a point mass.  This means that the 
rate of stellar evaporation across the tidal boundary is not strictly correct.  Regardless, 
based on previous studies that compared the rate of stellar evaporation in MOCCA to 
that in $N$-body simulations, which incorporate a more realistic treatment of the Galactic 
tidal field \citep[e.g.][]{giersz08,heggie08}, this did not significantly affect our 
results.  Additionally, MOCCA cannot handle time-dependent 
tidal fields, as is the case for GCs on eccentric orbits (although 
Sollima \& Mastrobuono Battisti 2014 recently proposed a method to address 
this issue within the Monte Carlo approach).  Thus, our results are based 
on simulations of GCs placed on purely circular orbits, since MOCCA \textit{can} treat 
a static tidal field with comparable accuracy to currently available $N$-body codes 
\citep{fukushige00}.  

There are two primary sources of stochasticity in the MC method.  The first relates to the 
determination of the position of an object along a given orbit.  At each time-step, the 
position is determined according to the probability of finding an object in a particular 
location along its orbit.  At each subsequent time-step, a new position is determined, and 
this can differ substantially from the previous position.  This leads to stochastic changes 
in the gravitational potential between time-steps.  The second source of stochasticity pertains 
to the outcomes of dynamical encounters involving 
binaries.  The initial conditions going into the dynamical encounters (whose outcomes are 
calculated using the FEWBODY code) are also mildly stochastic, due to the aforementioned stochastic 
changes to the positions of objects between time-steps.  This translates into some stochasticity 
in the outcomes of the interactions, since these depend sensitively on the initial interaction 
conditions.  

We do not expect either of these sources of stochasticity to have significantly affected our results.  
We tested this by re-running the models with the same initial conditions, but using 
different random seed values for the random number generators.  Our results always yield the same 
final cluster parameters (half-mass radius, density, mass, etc.) to within typically a few percent.  
The only exception to this is the core radius, which fluctuates significantly (by as much as a factor 
of a few) from time-step to time-step.  Importantly, however, this is the case not only for 
the Monte Carlo approach, but also for $N$-body models for star cluster evolution \citep[e.g.][]{webb14}.  

Other limitations of the Monte Carlo method include deviations from spherical symmetry including 
rotation (however Vasiliev recently proposed a method to circumvent this issue using the 
Monte Carlo method; private communication), sudden changes in the gravitational potential 
(e.g. violent relaxation or sudden gas 
removal) and hierarchical multiples (e.g. encounters involving triple stars, quadruples, etc.).  Most 
of these issues are currently being investigated, and should be included in the MOCCA code in the near 
future.  The Monte Carlo method also requires that the local time-step is always a fraction of the 
local relaxation time, which must always be longer than the local dynamical time of the system.  This 
condition is always satisfied in our models, particularly during the first few Myr of cluster evolution 
when most (soft) primordial binaries are disrupted due to dynamical encounters. 

\section{Summary} \label{summary}

In this paper, we consider the origins of the binary properties observed in Galactic 
globular clusters.  To this end, we present the results of a suite of Monte Carlo models 
for GC evolution performed using the MOCCA code \citep{giersz13}, which we compare to the 
observed present-day binary fractions of \citet{milone12}.  

In Paper I, we showed that the observed 
distribution of present-day \textit{central} binary fractions can be reproduced assuming an 
universal initial binary fraction of 10\%, a period distribution flat in the logarithm of 
the binary orbital semi-major axis and moderate initial densities close to the present-day 
values observed in Galactic GCs ($\sim$ 10$^2$-10$^3$ M$_{\odot}$ pc$^{-3}$).  
In this paper, we consider instead an universal initial binary fraction near unity, the binary 
orbital parameter distributions of \citet{kroupa95} (i.e. with a significant soft binary 
component) and high initial densities 
(10$^4$-10$^6$ M$_{\odot}$ pc$^{-3}$).  Our results suggest that, to first-order, the 
present-day binary fractions inside the half-mass radius are degenerate.  That is, they can 
be reproduced assuming either initially low binary fractions with a dominant hard component and 
moderate densities, or initially high binary fractions with a dominant soft component and high 
densities.  We show that the observed present-day binary fractions 
\textit{outside} the half-mass radius can break this degeneracy.  In this regard, our results 
are the most consistent with high initial binary fractions and high initial densities, since 
these conditions are needed to reproduce the observed anti-correlation between the total cluster 
mass and the observed binary fractions outside r$_{\rm h}$.  To reproduce the slope of this 
anti-correlation, our results favour an initial mass-density relation 
r$_{\rm h}$ $\propto$ M$_{\rm clus}^{\alpha}$ with $\alpha < 1/3$.  This corresponds to initially 
more massive clusters having higher initial densities, in rough agreement with the results of 
Paper I of this series \citep{leigh13c}.  Additionally, the observed present-day
mean densities inside r$_{\rm h}$ are typically in the range 
$\sim$ 10$^2$-10$^4$ M$_{\odot}$ pc$^{-3}$ for Galactic GCs \citep{harris96}, which 
agree better with our initially tidally-underfilling models.  

Thus, we require high initial densities 
($\sim$ 10$^4$-10$^6$ M$_{\odot}$) with initially tidally-underfilling clusters to reproduce the 
\textit{observed} anti-correlation between 
binary fraction and cluster mass outside r$_{\rm h}$, in conjunction with present-day densities 
that agree with the observed range in Galactic GCs.  Importantly, such high densities have 
actually been observed in young massive star-forming regions 
\citep[e.g.][]{hillenbrand98,kuhn14}.

We further apply the extrapolation technique first introduced in \citet{leigh12} to 
the initial and final binary orbital energy distributions, in order to quantify the 
degree of dynamical processing as a function of the assumed initial conditions.  We 
illustrate that our method can be used to constrain both the initial cluster density as well as 
the initial mass-density relation, if/when more sophisticated observations become available 
for the underlying present-day binary orbital parameter distributions in GCs.  We have 
only applied our technique using two different initial mass-density relations (i.e. the 
tidally-filling and tidally-underfilling sets of models) and a single universal set 
of binary fractions and orbital parameter distributions, using these as a ``proof-of-concept'' 
for our method and to highlight some of the effects of the assumed initial cluster density.  In 
principle, however, our method can be applied 
to \textit{any} choice of initial mass-density relation combined with \textit{any} choice 
of initial binary fractions and orbital parameter distributions (assuming they are either 
universal or show a clear and direct dependence on the initial mass-density relation).  This 
offers the potential for future studies to generate an extensive set of simulated old ($\sim$ 12 Gyr) 
GCs spanning three decades in present-day total cluster mass with minimal computational expense.  

As an overall conclusion, the results of this work are consistent with primordial GCs having 
formed with universal stellar populations in terms of both the initial mass function and the 
initial binary properties for stars less massive than $\sim$ 1 M$_{\odot}$.  For at least 
this stellar mass range, there is no evidence 
to suggest that the initial binary populations were significantly different a Hubble time ago in 
young massive clusters compared to what is observed today in nearby, young and sparse 
star-forming regions in the Milky Way disk.

%

\section*{Acknowledgments}

This work was partly supported by the Polish Ministry of Science and Higher 
Education, and by the National Science Centre
through the grants DEC-2012/07/B/ST9/04412 and DEC-2011/01/N/ST9/06000.  NL, 
JW, AS and COH all gratefully acknowledge the support of NSERC.  MM was partly 
supported through DFG grant KR 1635/40-1.  COH also 
acknowledges the support of an Alberta Ingenuity New Faculty Award.


\bsp

\label{lastpage}

\end{document}